\documentclass[twocolumn,pra,superscriptaddress,showkeys,amsmath]{revtex4-1}		
\usepackage{graphicx}
\usepackage{color}
\usepackage{hyperref}% add hypertext capabilities

\graphicspath{{figures_manuscript/}}
\usepackage{verbatim}
\usepackage[normalem]{ulem} 

\begin{document} 

\title{Time-dependent spectra of a three-level atom in the presence of electron 
shelving}  
 
\author{R. Rom\'an-Ancheyta}
\affiliation{Instituto de Ciencias F\'isicas, Universidad Nacional Aut\'onoma 
  de M\'exico, Apartado Postal 48-3, 62251 Cuernavaca, Morelos, M\'exico}
\affiliation{Department of Physics, Ko\c{c} University, \.{I}stanbul, Sar{\i}yer, 
	34450, Turkey}	
\author{O. de los Santos-S\'anchez} 	
\affiliation{Instituto de Ciencias F\'isicas, Universidad Nacional Aut\'onoma 
	de M\'exico, Apartado Postal 48-3, 62251 Cuernavaca, Morelos, M\'exico}
\author{L. Horvath}
\affiliation{Department of Physics and Astronomy, Macquarie University, 
	Sydney, NSW 2109, Australia}
\author{H. M. Castro-Beltr\'an} 
\email{hcastro@uaem.mx}
\affiliation{Centro de Investigaci\'on en Ingenier\'ia y Ciencias 
Aplicadas and Instituto de Investigaci\'on en Ciencias B\'asicas y Aplicadas, \\ Universidad Aut\'onoma del Estado de Morelos, 
Avenida Universidad 1001, 62209 Cuernavaca, Morelos, M\'exico}

%\draft 
\date{\today}
\begin{abstract}
We investigate time-dependent spectra of the intermittent resonance 
fluorescence of a single, laser-driven, three-level atom due to electron 
shelving. After a quasi-stationary state of the strong transition, a slow 
decay due to shelving leads to the steady state of the three-level system. 
The long-term stationary spectrum consists of a coherent peak, an 
incoherent Mollow-like structure, and a very narrow incoherent peak at 
the laser frequency. We find that in the ensemble average dynamics the 
narrow peak emerges during the slow decay regime, after the Mollow 
spectrum has stabilized, but well before an average dark time has passed. 
The coherent peak, being a steady state feature, is absent during the time 
evolution of the spectrum. 
\end{abstract}

%\pacs{42.50.Lc, 42.50.Md, 32.80.-t}
\maketitle

%%%%%%%%10%%%%%%%%20%%%%%%%%30%%%%%%%%40%%%%%\section{Introduction}  
\textit{Introduction.---} Electron shelving occurs in atoms when the stream 
of photons emitted by a laser-driven strong transition is interrupted by 
quantum jumps to metastable states; these jumps introduce finite dark 
periods, hence blinking, in the resonance fluorescence scattering. The 
blinking or intermittency of the fluorescence is a stationary random process 
whose statistics of bright and dark periods are well studied \cite{NaSD86,SNBT86,BHIW86,PlKn98}. Recently, it was shown to be 
possible to reverse the onset of a dark period \cite{Minev18}. The photon 
statistics \cite{MeSc90} and phase-dependent fluctuations \cite{CaRG16} 
of blinking resonance fluorescence have also been studied in some detail. 

The atom's ensemble average resonance fluorescence shows signatures 
of shelving. The population of the excited state of the strong transition, for 
example, reaches a short term quasi-stationary state (typical of the 
two-level system) followed by a long decay to the final steady state at nearly 
the decay rate of the weak transition \cite{PlKn98}. Stationary spectra of 
blinking resonance fluorescence have also been studied: Hegerfeldt and 
Plenio \cite{HePl95} and Garraway et al \cite{GaKK95} found that for a 
bichromatically driven V- and $\Lambda$-type three-level atom (3LA) the 
spectrum consists of a delta-peaked coherent term, an incoherent 
Mollow-like spectrum \cite{Mollow69}, and a novel feature given by a 
narrow inelastic peak. This narrow peak is the spectral signature of the slow 
decay of the atomic populations, caused by the presence of a slow decay 
channel that randomly interrupts the fluorescence of a strongly driven 
transition. The narrow peak was measured by B\"uhner and Tamm with a 
single $^{171} \mathrm{Yb}^+$ ion by heterodyne detection \cite{BuTa00}. 
Evers and Keitel \cite{EvKe02} then proved that the narrow peak grows 
at the expense of the coherent peak, as the difference between the 
intensity of the coherent peaks of a two-level atom (2LA) and a 3LA. 

Little attention has been paid to the spectrum of blinking resonance 
fluorescence as a dynamical observable. Only the spectrum during a single 
bright period, of variable length, has been considered so far \cite{HePl96}; 
it was the Mollow spectrum, proving that the narrow peak is a feature of the 
random interruption of the fluorescence. One then asks how the narrow peak 
emerges if the dark periods are taken into account during the ensemble 
average measurement of the spectrum.  

In this paper we investigate time-dependent spectra of a single three-level 
atom undergoing blinking resonance fluorescence, that is, including both 
bright and dark periods in the ensemble evolution. Our main result is that 
the narrow inelastic component due to electron shelving develops much 
later than the two-level Mollow spectrum, but before the average dark time 
has passed. 

For this purpose we calculate the Eberly-W\'odkiewicz (EW) physical 
spectrum \cite{EbWo77}, which gives the most rigorous theoretical 
description for time-dependent spectra. In this model, the source field is 
scanned by a nonzero bandwidth filter prior to photodetection, handling 
properly the time-energy uncertainty that arises when both time and 
frequency are to be resolved. The EW spectrum has been applied to study 
nontrivial dynamics of optical systems, for example: the effects of 
switching-on \cite{EbKW80} and switching-off the laser \cite{HuTE82}, initial 
atomic coherence \cite{GoMo87}, and coherent population trapping 
\cite{JLDS89} in resonance fluorescence; spontaneous emission (the first 
prediction of the Rabi doublet) \cite{SaNE83}, Dicke superradiance 
\cite{CaSC96} and frequency-filtered photon correlations \cite{Valle12} in 
cavity QED. The EW spectrum has also been applied to the spontaneous 
emission in front of a moving mirror \cite{GHD+10,Mirza15} and two-atom 
entanglement \cite{HoFi10} in QED.

%%%%%%%%10%%%%%%%%20%%%%%%%%30%%%%%%%%40%
%\section{Model}
%
\begin{figure}[h]
\includegraphics[width=3.7cm,height=3.5cm]{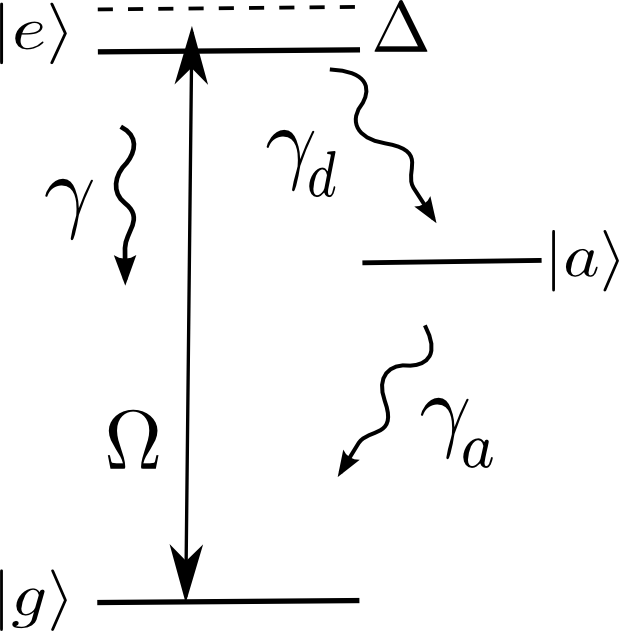} 
\caption{\label{fig:3LA} 
Scheme of the three-level atom showing laser excitation of the 
$|e \rangle - |g \rangle$ transition with Rabi frequency $\Omega$, detuning 
$\Delta$, and spontaneous decay rate $\gamma$, and spontaneous 
decay via the metastable state $| a \rangle$ at rates 
$ \gamma_d, \gamma_a$. } 
\end{figure}
\textit{Model.---} Our system, depicted in Fig.~\ref{fig:3LA}, consists of a 
three-level atom with one laser-driven transition with Rabi frequency 
$\Omega$, detuning $\Delta$ and decay rate $\gamma$, whose 
fluorescence is monitored. The excited state $|e \rangle$ also decays to 
a long-lived intermediate state $|a \rangle$ at the rate $\gamma_d$, and 
from this to the ground state at the rate $\gamma_a$. 

The Markovian master equation in the frame rotating at the laser frequency is 
\begin{eqnarray}
\dot{\rho} &=& -i [\mathcal{H},\rho] +{\gamma}\mathcal{L}[\sigma_{ge}] \rho 
	+{\gamma_d} \mathcal{L}[\sigma_{ae}] \rho 	
	+ {\gamma_a} \mathcal{L}[\sigma_{ga}] \rho   , 
\end{eqnarray}
where $\mathcal{H} = \Delta \, \sigma_{eg} \sigma_{ge} 
	+\Omega ( \sigma_{eg} +\sigma_{ge})/2$ 
is the atom-laser Hamiltonian in the rotating wave approximation and 
$\mathcal{L}[\mathcal{O}]\rho \equiv \mathcal{O}\rho\mathcal{O}^\dagger 
 - (\mathcal{O}^\dagger\mathcal{O}\rho 
+\rho\mathcal{O}^\dagger\mathcal{O})/2$ 
are spontaneous decay superoperators. The atomic operators 
$\sigma_{jk} = |j \rangle\langle k|$ obey 
$\sigma_{jk} \sigma_{lm} = \sigma_{jm} \delta_{kl}$.  

Because of the pure spontaneous emission decay, the incoherent nature of 
the $|e \rangle - |a \rangle - |g \rangle$ channel decouples the equations for 
the coherences involving the $|a \rangle$ state from those of the laser 
driven $|e \rangle - |g \rangle$ transition \cite{CaRG16,EvKe02}. The Bloch 
equations of the effective two-level system can then be written in compact 
form as 
\begin{eqnarray} 	\label{eq:BlochEqs}
\langle \dot{\mathbf{s}} (t) \rangle 
	&=& \mathbf{M} \langle \mathbf{s}(t) \rangle +\mathbf{b} 	, \\ 
\mathbf{s} &\equiv&   
	\left( \sigma_{ge},  \sigma_{eg}, \sigma_{ee}, \sigma_{gg}  \right)^T , \\ 
\mathbf{b} &=& (0,0,0,\gamma_a)^T 	,  
\end{eqnarray}
\begin{eqnarray} 	\label{eq:matrixM}
\mathbf{M} &=& \left( \begin{array}{cccc}
-i \Delta -\gamma_+/2 & 0 & i\Omega/2 & -i\Omega/2 \\ 
0 & i \Delta -\gamma_+/2 & -i\Omega/2 & i\Omega/2 	 \\ 
i\Omega/2 & -i\Omega/2 & - \gamma_+  & 0 		\\ 
-i\Omega/2 & i\Omega/2 & \gamma_- & -\gamma_a 
\end{array}  \right)    ,	\nonumber \\
\end{eqnarray}
\begin{eqnarray} 
\gamma_+ = \gamma + \gamma_d   , 	\qquad 
\gamma_- = \gamma - \gamma_a . 
\end{eqnarray}
Above, $\dot{\mathbf{s}}$ is the derivative of $\mathbf{s}$ with respect to 
time. 

In general, the Bloch equations are solved numerically. However, accurate 
approximate analytical solutions in the resonant case, $\Delta =0$, in the 
regime (\ref{eq:shelvingCond}) were obtained by two of us in \cite{CaRG16}. 
The populations and coherences show the typical short-term decay at the 
rate $3\gamma_+/4$ reminiscent of the 2LA dynamics and a long-term 
decay, at roughly $\gamma_a$, that signals shelving in the metastable state 
$|a \rangle$ \cite{PlKn98}.  

The solutions in the steady state are  
\begin{subequations}
\begin{eqnarray}
\langle \sigma_{eg} \rangle_{st} 
	&=& \frac{i \Omega[ \gamma_+ + i 2\Delta ]}
	{(2 +q)\Omega^2 +\gamma_+^2 +4\Delta^2} ,  \\ 
\langle \sigma_{gg} \rangle_{st} 
	&=& \frac{\Omega^2 +\gamma_+^2 +4\Delta^2}
	{(2 +q)\Omega^2 +\gamma_+^2 +4\Delta^2}   , \\ 
\langle \sigma_{ee} \rangle_{st} &=& \frac{\Omega^2}
	{(2 +q)\Omega^2 +\gamma_+^2 +4\Delta^2}   , 
	\label{eq:rho_eeSS}	\\ 
\langle \sigma_{aa} \rangle_{st} &=& \frac{q \Omega^2}
	{(2 +q)\Omega^2 +\gamma_+^2 +4\Delta^2}    , 	 
\end{eqnarray}
\end{subequations}
where  
\begin{eqnarray}
q &=& \gamma_d / \gamma_a   . 
\end{eqnarray}
and 
$\langle \sigma_{ge} \rangle_{st} = \langle \sigma_{eg} \rangle_{st}^{\ast}$.

This system features blinking, with long bright and dark periods in the 
fluorescence of the $|e \rangle - |g \rangle$ transition due to electron 
shelving in the metastable state $| a \rangle$, if the decay rates obey the 
relation  
\begin{eqnarray} 	\label{eq:shelvingCond}
\gamma \gg \gamma_d , \gamma_a   .  
\end{eqnarray}
A random telegraph model can be used to calculate the average length 
of the bright and dark periods \cite{EvKe02,PeKn88}. For this  derivation 
the equation for the metastable state, 
$\dot{\rho}_{aa} = \gamma_d \rho_{ee} - \gamma_a \rho_{aa}$, is needed 
($\rho_{jk} = \langle \sigma_{kj} \rangle$). During a bright period the state 
$|a \rangle$ is never occupied, $\rho_{aa}(t) =0$. The average bright time 
$T_B$ is defined as $T_B^{-1} = (\dot{\rho}_{aa})_{t \to \infty}$, where the 
limit means a time long enough for the two-level transition 
$|g \rangle - |e \rangle$ to reach the steady state, so 
$\rho_{ee}(\infty) \to (\rho_{ee}^{st})_{2LA}$. Thus, with $q=0$ and 
$\gamma_+ \to \gamma$ in Eq.~(\ref{eq:rho_eeSS}), we have   
\begin{eqnarray}    \label{eq:AvBrightPeriod}
T_B = \frac{2\Omega^2 +\gamma^2 +4\Delta^2}
  	{\gamma_d \Omega^2} . 		
\end{eqnarray}
Similarly, the average dark time $T_D$ is defined as 
$T_D^{-1} = (\dot{\rho}_{aa})_{t \to \infty}$ but, during a dark period 
$\rho_{aa}(t) =1$ and $\rho_{ee}(t) = 0$, hence  
\begin{eqnarray}    \label{eq:AvDarkPeriod}
	T_D = \gamma_a^{-1} 	.
\end{eqnarray}

The three-level scheme of Fig.~\ref{fig:3LA} is a simplified theoretical 
representation of the complex energy level structure of an 
$^{171} \mathrm{Yb}^+$ ion under the driving configuration presented in 
\cite{BuTa00}. In this paper the stationary spectrum of 
$^{171} \mathrm{Yb}^+$ was measured where, in order to reduce the dark 
periods in the ion's fluorescence, additional incoherent pumping from 
$| a \rangle$ to a fourth level (not shown) with faster decay to $| g \rangle$ 
was applied. Thus, $\gamma_d$ is considered an effective decay rate that 
includes such pumping. 

%%%%%%%%10%%%%%%%%20%%%%%%%%30%%%%%%%%40%%
%\section{Stationary Power Spectrum} 
\textit{Stationary Power Spectrum.---} The stationary Wiener-Khintchine 
power spectrum is given by the Fourier transform of the field autocorrelation 
function \cite{FiTa17},  
\begin{eqnarray}
S(\omega) &=& \mathrm{Re} \int_0^{\infty} d\tau e^{-i \omega \tau} 
	\langle \sigma_{eg} (0) \sigma_{ge} (\tau) \rangle_{st}  .
\end{eqnarray}
By writing the atomic operators as the sum of a mean, 
$\langle \sigma_{jk} \rangle_{st}$, plus fluctuations, 
$\tilde{\Delta} \sigma_{jk}(t)$, that is,  $\sigma_{jk} (t) 
= \langle \sigma_{jk}  \rangle_{st}  +\tilde{\Delta} \sigma_{jk} (t) $, 
we can separate the spectrum into a coherent part 
\begin{eqnarray} 	\label{eq:I_coh}
S_{coh}(\omega) &=& |\langle \sigma_{eg} \rangle_{st} |^2 
	\mathrm{Re} \int_0^{\infty}  e^{-i \omega \tau} d\tau 
	= \pi |\langle \sigma_{eg} \rangle_{st} |^2  \delta(\omega)	, \nonumber \\ 
&=& \frac{ \pi \Omega^2 (\gamma_+^2 +4\Delta^2) }
    {[(2+q)\Omega^2 +\gamma_+^2 +4\Delta^2 ]^2 } \delta(\omega)	   , 
\end{eqnarray}
due to elastic scattering, and an incoherent part 
\begin{eqnarray} 	\label{eq:Sinc}
S_{inc}(\omega) &=&  \mathrm{Re} \int_0^{\infty} d\tau e^{-i \omega \tau} 
	\langle \tilde{\Delta} \sigma_{eg}(0) 
	\tilde{\Delta} \sigma_{ge}(\tau) \rangle_{st} , 	
\end{eqnarray}
due to atomic fluctuations. For the strong transition of the V and $\Lambda$ 
3LA's, $S_{inc}(\omega)$ consists of a spectrum nearly identical to the 2LA 
Mollow one (peaks of width of the order of $\gamma$, a single one in the 
weak driving limit and a triplet in the strong excitation regime \cite{Mollow69}) 
plus a narrow peak of nearly Lorentzian shape at the laser frequency due to 
the presence of electron shelving \cite{HePl95,GaKK95}.  

B\"uhner and Tamm experimentally measured the narrow peak near the 
saturation regime by heterodyne detection \cite{BuTa00}. Later, Evers and 
Keitel \cite{EvKe02} studied the narrow peak in detail and found that it 
comes at the expense of the coherent peak of the 2LA spectrum. Noting in Eq.~(\ref{eq:I_coh}) that $q >0$, the coherent peak of the 3LA is smaller 
than that of the 2LA. Writing 
$\left(S_{coh} \right)_{NLA} =I_{NLA} \delta(\omega)$, for $N=2,3$, the 
relative intensity of the narrow inelastic peak is given by the difference in 
the size of the coherent peak of the two- and three-level atoms, 
$I_{np} = I_{2LA} - I_{3LA}$, 
\begin{eqnarray} 	\label{eq:I_narrowpeak}
I_{np} &=& \left( |\langle \sigma_{eg} \rangle_{st} |^2 \right)_{2LA}
	- \left( |\langle \sigma_{eg} \rangle_{st} |^2 \right)_{3LA} 	\nonumber \\ 
&=& \frac{ \Omega^4 [(2+q) \gamma^2 -2\gamma_+^2 +4\Delta^2 q ]}
	{[2\Omega^2 +\gamma^2 +4\Delta^2 ]^2
	[(2+q)\Omega^2 +\gamma_+^2 +4\Delta^2 ]^2 }  . \nonumber 	\\
\end{eqnarray}
The narrow peak becomes smaller for increasing Rabi frequencies, but 
increasing detuning enhances the peak if the Rabi frequency is 
increased \cite{EvKe02}; this peak is the largest for a detuning 
$\Delta_{max}^2 = \left[ (q-2) \Omega^2 -2 \gamma^2 \right]/8$. The width 
of the narrow peak is accurately given by \cite{HePl95,EvKe02} 
\begin{eqnarray} 	\label{eq:widthextra}
\Gamma_{np} &=& T_D^{-1} +T_B^{-1} 	\nonumber \\
	&=& \gamma_a \left[ 1+ \frac{ q\Omega^2 } 	
	{2\Omega^2 +\gamma^2 +4\Delta^2}  \right]     .  	  
\end{eqnarray}
An analytic formula for the full stationary spectrum on resonance in the 
regime (\ref{eq:shelvingCond}) has been given in \cite{CaRG16}. 

%%%%%%%%10%%%%%%%%20%%%%%%%%30%%%%%%%%4
%\section{Time-Dependent Spectrum}
\textit{Time-Dependent Spectrum.---} We calculate time-dependent spectra 
(TDS) using the physical spectrum of Eberly and W\'odkiewicz 
\cite{EbWo77}     
\begin{eqnarray}
S(D,t,\Gamma) &=& \Gamma \int_{t_0}^t dt_1 
 \int_{t_0}^t dt_2 \,\, e^{-(\Gamma/2 -i D)(t-t_1)} \nonumber 	\\
&& \times	e^{-(\Gamma/2 +i D)(t-t_2)} 
	\langle \sigma_{eg}(t_1) \sigma_{ge}(t_2) \rangle   , 
\end{eqnarray}
where $D=\omega - \omega_l$ is the detuning of the laser frequency 
$\omega_l$ from the filter's frequency $\omega$, and $\Gamma$ is the 
filter's bandwidth. Admittedly, the calculation of TDS is not a simple task, 
and more often than not a numerical solution is required. Some authors 
often wish to avoid the filter effects and resort to simpler, yet probably 
defective, approaches \cite{EbWo77,FiTa17}. The inclusion of the filter 
ensures that the time-energy uncertainty is properly accounted for in 
theoretical calculations. An additional benefit of filtering is that it can 
enhance important features and the signal to noise ratio in the measured 
TDS of weak signals. 

For computation purposes it is convenient to rewrite the double integral in 
terms of integrals for $t_2$ and $\tau = t_1 -t_2$ \cite{EbKW80}; making 
$t_0 =0$ we have 
\begin{eqnarray}
S(\omega, t, \Gamma) &=& 2\Gamma \mathrm{Re} \Big{[}
\int_{0}^t dt_2  e^{-\Gamma(t-t_2)}\int_{0}^{t-t_2} d\tau \, 
	e^{(\Gamma/2 -i D)\tau} 	\nonumber 	\\
&& \times \langle \sigma_{eg}(t_2+\tau)\sigma_{ge}(t_2)\rangle  \Big{]} .
\end{eqnarray}

To solve for the two-time correlations we apply the quantum regression 
formula \cite{Carm02} to Eq.~(\ref{eq:BlochEqs}) obtaining  
\begin{eqnarray}
\partial_{\tau} \langle \mathbf{u}(t_2,\tau) \rangle 
	&=& \mathbf{M} \langle \mathbf{u} (t_2,\tau) \rangle +\mathbf{c} (t_2) , 
\end{eqnarray}
where
\begin{eqnarray*}
\mathbf{u} (t_2,\tau) &=&   
	\big[  \sigma_{ge} (t_2+\tau) \sigma_{ge} (t_2) ,  
	\sigma_{eg} (t_2+\tau) \sigma_{ge} (t_2) , 		 	\\
	&&  \sigma_{ee} (t_2+\tau) \sigma_{ge} (t_2) , 
	\sigma_{gg} (t_2+\tau) \sigma_{ge} (t_2) \big]^T  , \\ 
\mathbf{c}(t_2) &=& (0,0,0, \gamma_a \langle \sigma_{ge} (t_2) \rangle)^T  ,  			\nonumber 
\end{eqnarray*}
which we solve numerically with initial condition 
$\mathbf{u}(t_2,0) = \left(0, \sigma_{ee} (t_2),  0, \sigma_{ge} (t_2) \right)^T$.
The number of parameters in our system makes it very difficult to obtain 
analytical expresions for the TDS.   

\begin{figure}[b]
\includegraphics[width=8.5cm,height=5.5cm]{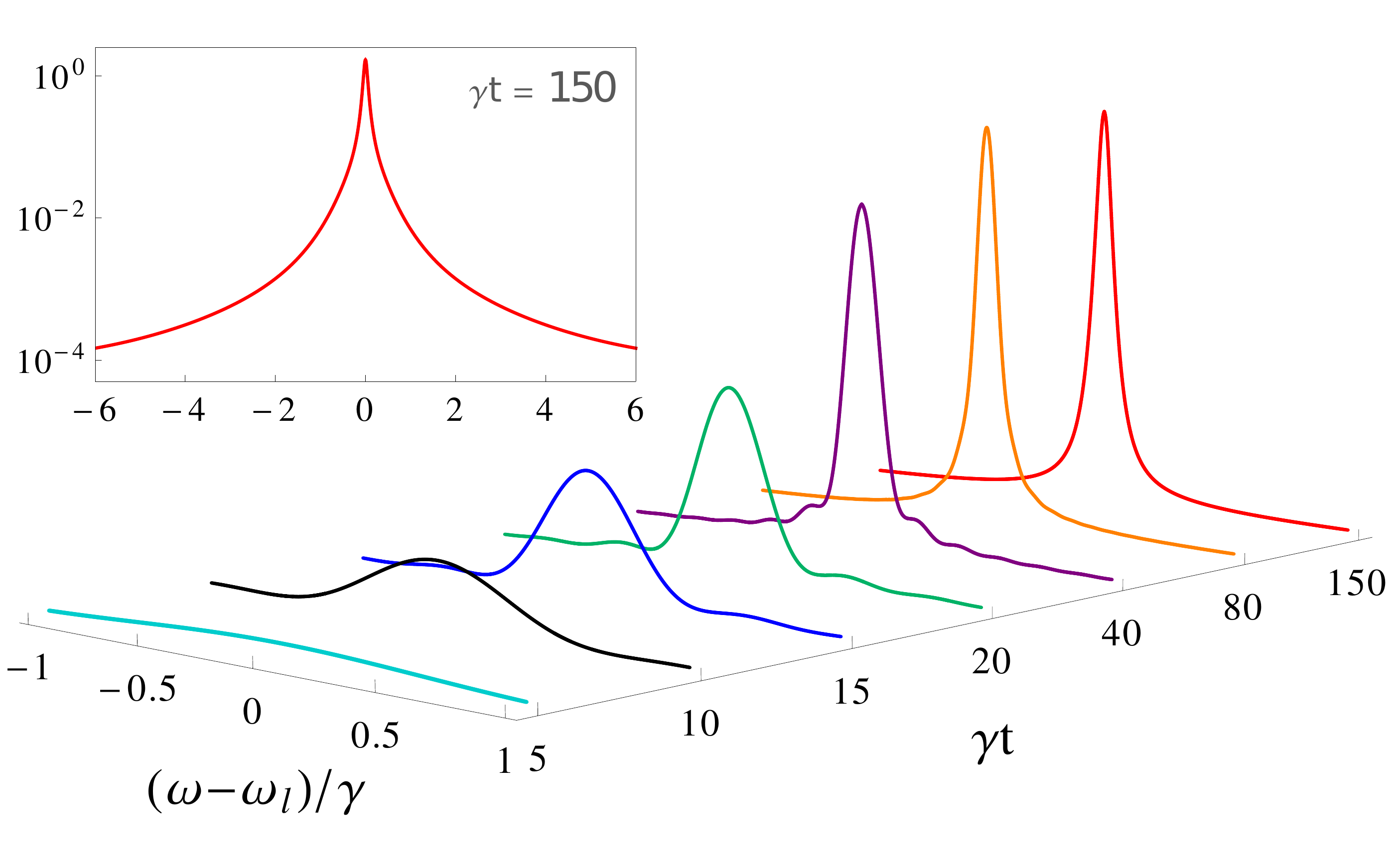} 
\caption{\label{fig2:tds_sat} 
Time-dependent spectra for moderate laser field strength, 
$\Omega =\gamma_+/4 =0.2625 \gamma$, $\gamma_d=0.05 \gamma$ 
and $\gamma_a=0.015 \gamma$. The filter's bandwidth is 
$\Gamma = 0.1 \gamma$. The inset shows the spectrum at $\gamma t =150$ 
in semi-log scale and wider frequency range to reveal the broad 
component. }
\end{figure}

\begin{figure}[t]
\includegraphics[width=8.5cm, height=5.1cm]{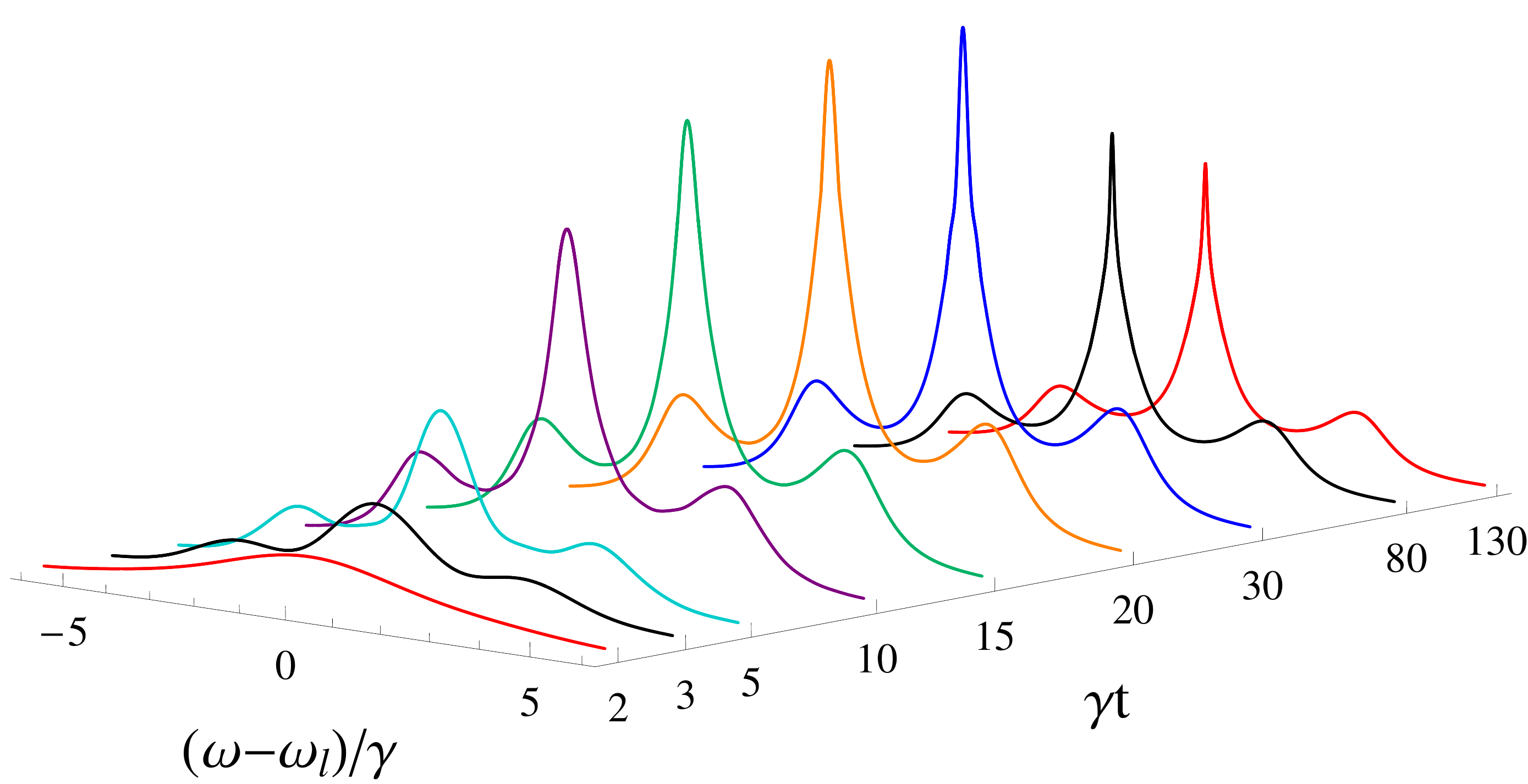} 
\caption{\label{fig3:tds_strong} 
Same as Fig.~\ref{fig2:tds_sat} but for strong driving, 
$\Omega=3.5\gamma$.}
\end{figure}

\begin{figure}[b]
\includegraphics[width=8.5cm, height=5.5cm]{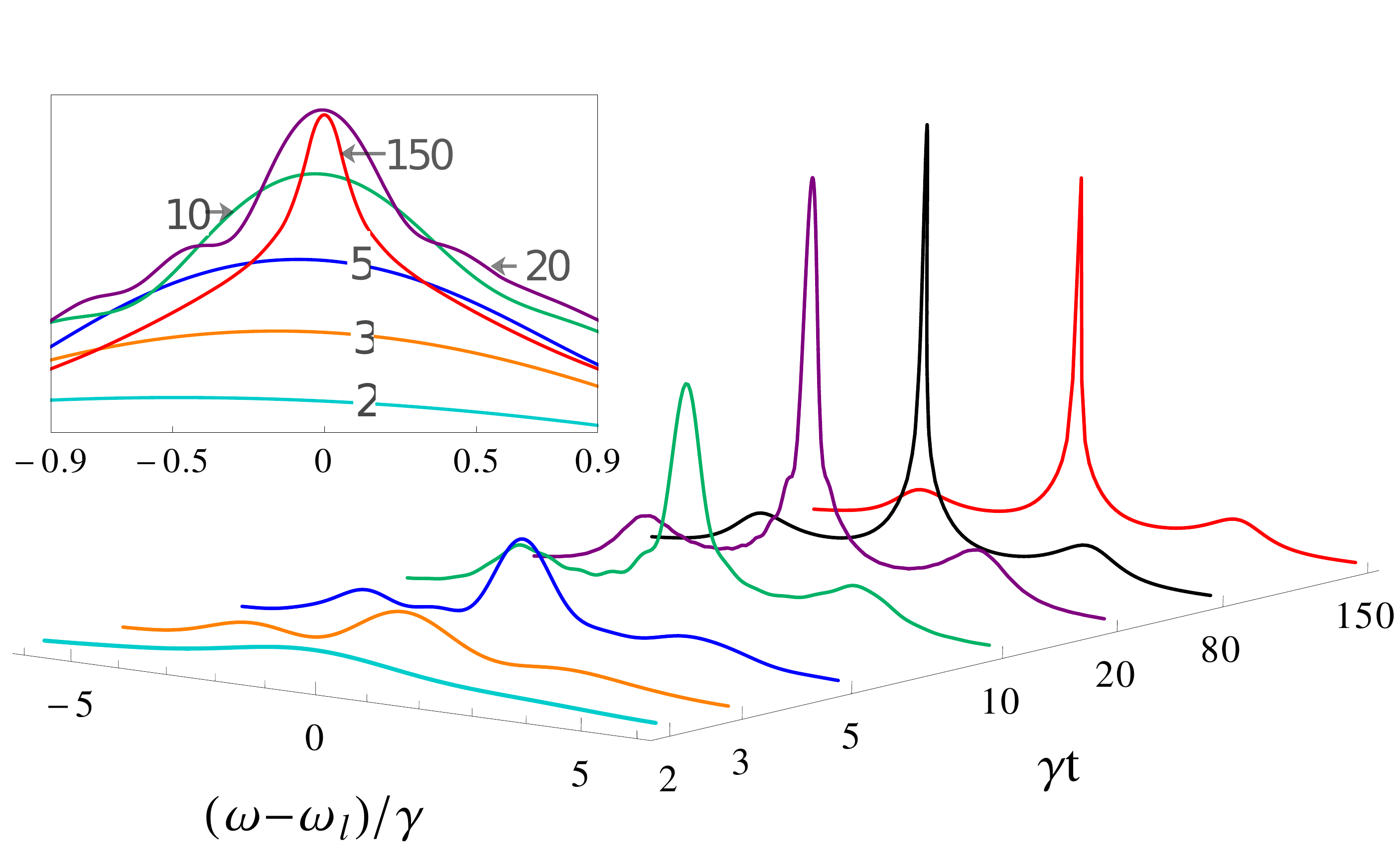} 
\caption{ 
TDS for strong field, $\Omega=3.5\gamma$, but detuning $\Delta=1$. 
The other parameters as in Fig.~\ref{fig2:tds_sat}. The inset shows the 
diminishing asymmetry of the center of the spectrum for increasing time. }
\label{fig4:tds_strong_det}
\end{figure}

Figures~\ref{fig2:tds_sat}-\ref{fig4:tds_strong_det} show our results for the 
TDS of our blinking system. Figure~\ref{fig2:tds_sat} displays the spectra 
in the excitation regime near saturation, $\Omega = \gamma_+/4$. A narrow 
peak develops for long times, $\gamma t \gg 1$, above a background given 
by the usual broad peak of width $\sim \gamma$ formed on a shorter time 
scale of several lifetimes, $\gamma^{-1}$. 

To better appreciate the different time scales for the appearance of the 
spectral components we show the TDS in the strong field regime, 
$\Omega = 3.5 \gamma$. In Fig.~\ref{fig3:tds_strong}, while the triplet is 
well developed for times $\gamma t \sim 10$ the narrow peak arises at 
about $\gamma t \sim 20$. As expected from the stationary spectrum, the 
narrow peak in the strong field regime is smaller than in the saturation 
regime \cite{EvKe02,CaRG16}. Hence, as suggested in \cite{EvKe02}, 
some detuning notably enhances the narrow peak against the spectral 
background of the Mollow triplet, as shown in Fig.~\ref{fig4:tds_strong_det}. 
A slight asymmetry occurs in the detuned case that vanishes in the long 
time limit \cite{EbKW80}; in this case one of the sidebands is closer to the 
atomic resonance and is larger than the other \cite{EbKW80}, while the 
asymmetry in the center of the spectrum gets smaller (see inset). More 
pronounced spectral asymmetries are found, for example, in detuned 
pulsed laser resonance fluorescence \cite{GuMH18}. 

It is important to note that while the narrow peak develops much later than 
the Mollow spectrum, it does actually emerge, if not stabilize, well before 
an average dark time $T_D$ has passed. The presence of dark periods in 
the fluorescence is felt soon in the ensemble's evolution: in some 
realizations of the ensemble the dark period may occur before the bright 
one. From Eqs.~(\ref{eq:AvBrightPeriod}) and (\ref{eq:AvDarkPeriod}) it is 
seen that the average bright time depends on both laser and atomic 
parameters, while the average dark period depends only on the effective 
lifetime $\gamma_a^{-1}$ of the metastable state $|a \rangle$. In the TDS 
sequences of Figs.~\ref{fig2:tds_sat} -- \ref{fig4:tds_strong_det}, 
$\gamma T_D \simeq 67$, and $\gamma T_B \simeq 330, 42$, and 48, 
respectively. They reveal the time scale of the dark and bright periods in 
the ensemble evolution.   

We have to discuss also effects of the filter on the EW time-dependent 
spectrum. First, it could be argued that the observed narrow peak is the 
filter-broadened coherent spectral component. This is not the case because 
the delta peak is a steady state feature of the spectrum \cite{FiTa17}; it 
should not appear in a TDS, however long is the \textit{finite} observation 
time. What we undoubtedly see is the incoherent narrow peak produced by 
random interruptions in the fluorescence of the strong transition 
$|g \rangle - |e \rangle$ caused by the atom's excursions into the weak 
transition channel $|e \rangle \to |a \rangle \to |g \rangle$ \cite{HePl95}. 
Moreover, the narrow peak grows at the expense of the coherent peak 
\cite{EvKe02}: its intensity is the difference among the intensities of the 
coherent peak of the two- and three-level systems, 
Eq.~(\ref{eq:I_narrowpeak}).   

Another issue is the choice of filter bandwidth $\Gamma$. On one hand, 
it must be able to resolve the different spectral components, therefore 
$\Gamma$ should be a fraction of the width, $\sim \gamma$, of the Mollow 
spectral peaks. On the other, $\Gamma$ cannot be infinitely small, as is 
assumed for the stationary spectrum \cite{EbWo77}. The filter bandwidth 
in our plots, $\Gamma = 0.1 \gamma$, was chosen to focus on the narrow 
peak: for $\Gamma > \Gamma_{np}$ the filter sets the observed width of 
the narrow peak. 

The filter bandwidth also has dynamical consequences due to the 
time-energy uncertainty; the filter has to saturate in order to finish its 
transient effect and begin to produce stable spectra. This occurs after a 
time $\Gamma t >1$. Hence, a narrow filter $\Gamma < \gamma$ causes 
a delay in the stabilization of the fast-forming Mollow-like spectrum 
\cite{EbKW80}, while the narrow peak stabilizes soon since 
$\Gamma < \Gamma_{np}$. The transient effects on a spectrum are 
therefore felt for very long times, as seen in the temporary reduction of 
the spectra of Figs.~\ref{fig3:tds_strong} and \ref{fig4:tds_strong_det}. 
The different time scales due to atomic and filter parameters make it very 
difficult to fully assess the TDS analytically.  

Finally, we have used a density-operator-based approach, for which the 
TDS is the statistical average of infinitely many realizations. However, 
while the individual records of bright and dark periods are buried in the 
ensemble average, the impact of the latter on the TDS is evident in the 
emergence of the incoherent narrow peak. 

%%%%%%%%10%%%%%%%%20%%%%%%%%30%%%%%%%%40%%%
%\section{Conclusions}
\textit{Conclusions.---} We have investigated the time-dependent spectrum 
of intermittent resonance fluorescence and found that the narrow 
incoherent peak due to electron shelving emerges and stabilizes much later 
than the Mollow spectrum. We trust that an experimental observation of 
blinking resonance fluorescence TDS is within reach. TDS of two-level atom 
resonance fluorescence have been observed \cite{GoMo87} and 
measurements of shelving fluorescence have reached the accuracy required 
for applications such as precision measurements of fundamental constants 
and optical ion clocks \cite{GNJ+14,HLT+14}. We think that even for 
nonergodic blinking such as that of quantum dots or molecules \cite{StHB09}, 
whose TDS have been studied in \cite{LeBa15}, the Eberly-W\'odkiewicz 
physical spectrum would be of great benefit. The observation and 
interpretation of TDS could help to describe the dynamics of other systems 
with separate time scales such as super- and sub-radiance \cite{vLFL+13} 
and entanglement \cite{HoFi10} in collective atomic dynamics.

%%%%%%%%10%%%%%%%%20%%%%%%%%30%%%%%%%%40%%%%%%%%50%%%%%%%%60%%%%%%%%70%%%
\textit{Acknowledgments.---} R.~R.-A. wishes to thank CONACyT, Mexico 
for scholarship 379732. H.M.C.-B. thanks Prof. J. R\'ecamier for hospitality 
at ICF-UNAM.

%%%%%%%%%%%%%%%%%%%%%%%  REFERENCES %%%%%%%%
 %\end{thebibliography}

\end{document}